\documentstyle{elsart}
\input epsf
\begin{document}
\begin{frontmatter}
\title{Spiral Dynamics in Pattern-Forming Systems:
Mean Flow Effects}
\author{Lev S. Tsimring\thanksref{LT}}
\address{Institute for Nonlinear Science, University of California at San
Diego, La Jolla, CA 92093-0402}
\thanks[LT]{Supported by the U.S. DOE
under contracts DE-FG03-95ER14516 and DE-FG03-96ER14592}
\date{\today}
\begin{abstract}
Mean flow effects are discussed for two different pattern-forming
systems: Rayleigh-B\'{e}nard convection and Faraday instability in viscous 
fluid. In both systems spirals are observed in certain parameter regions. 
In the Rayleigh-B\'{e}nard convection, the spiral core instability and 
subsequent generation of up- and downflow hexagons are shown to occur 
due to the mean flow generated by the curved rolls near the core. 
In the Faraday instability, the mean flow which is generated by rapidly decaying
surface waves near the wall, causes wavenumber frustration which
leads to a rigid-body spiral rotation. In both cases we use phenomenological 
Swift-Hohenberg-type equations for the order parameter coupled to a large-scale 
mean flow. Numerical simulations are compared to recently reported experimental 
results.
\end{abstract}
\end{frontmatter}

\section{Introduction}
Nonlinear evolution of cellular patterns in extended systems near
the threshold of a primary instability are usually studied using
amplitude equations (e.g., Newell-Whitehead-Segel equation for 
roll patterns in isotropic systems\cite{NewWhit}). These amplitude equations 
can be derived from the first principles by perturbative
expansion using the supercriticality $\epsilon$ as a small parameter.
However, the expansion is based on a pre-chosen
orientation of the cellular pattern and thus do not preserve the 
full spatial symmetry of the physical system. Therefore this description fails 
for more complicated patterns such as coexisting domains with varying 
orientation of rolls, targets, spirals, etc. To overcome this difficulty,
phenomenological models similar to the so-called Swift-Hohenberg equation (SHE),
\begin{equation}
\partial\psi_t=\epsilon\psi-(1+\nabla^2)^2\psi-\psi^3
\label{she}
\end{equation}
are often employed (see \cite{CrHoh} and references therein). 
This model correctly describes the linear properties of the system
close to the threshold and yet preserves the rotational
symmetry of the system. The disadvantage of this approach is that 
the nonlinear term in this equation is usually added {\it ad hoc}
in order to provide a saturation of the primary instability and
cannot be systematically derived from underlying first
principles. Thus, while some of the features (including some secondary
instabilities, domain coarsening, etc.) of the pattern formation
can be modeled based on this equation, some other essential features
of the dynamics may be missing. In many examples these
features are related to the coupling between the cellular mode
and so-called zero mode slowly varying both in space and time. It
can be a concentration field in binary-fluid convection\cite{Riecke},
population inversion in extended lasers\cite{Moloney}, density field 
in granular media\cite{TA}, mean flow in 
Rayleigh-Benard convection\cite{SigZip,Mann,CrGr}. This problem can be
rectified by
expanding the basic model (\ref{she}) so that it includes the interaction 
with the zero mode. It allows to reproduce many subtle features of the 
dynamics of cellular patterns within a simple phenomenological description.

In this paper we illustrate these general ideas by two examples of spiral
dynamics in Rayleigh-B\'{e}nard convection and oscillating fluid layer. In
both cases the spiral rotation is intimately related to the interaction
with mean flow, self-generated by the cellular patterns in case of
convection, and wall-generated in case of Faraday system. We show that
recently observed spiral core instability\cite{PlBod,AssSt} and 
spiral-hexagon transition\cite{AssSt} in Rayleigh-B\'{e}nard convection 
in Boussinesq
fluid can be understood within the model for the order parameter coupled
with the mean flow\cite{AAST}. We also demonstrate that spiral rotation in Faraday
experiment can be explained by the wavenumber frustration which is
caused by the near-wall radial mean flow generated by decaying capillary
waves\cite{KKRT}.

\section{Spirals in Rayleigh-Benard convection}

Recently discovered spiral-defect chaos (SDC) in large aspect ratio
Rayleigh-Benard convection system at small Prandtl number\cite{SDC} was 
successfully reproduced both within Navier-Stokes equations\cite{Pesch} and 
also using SHE coupled to the equation for the mean flow 
generated by the curved rolls\cite{Vin},
\begin{eqnarray}
    \psi_t + ({\bf u} \cdot \nabla) \psi & = &
    \epsilon \psi - g \psi^3 + 3 (1 - g) (\nabla\psi)^2 \nabla^2 \psi -
    \nonumber \\
    & & - (1 + \nabla^2)^2 \psi \label{SHE} \\
    {\bf \Omega}_t - \sigma (\nabla^2 - c^2) {\bf \Omega} & = &
    g_m \hat{z} \cdot \nabla (\nabla^2 \psi) \times \nabla\psi
    \label{Omega} \\
    {\bf \Omega} & = & \nabla\times {\bf u}
    \label{vel}
\end{eqnarray}
Here $\psi$ is the order parameter, ${\bf u}$ the horizontal velocity
field of the large-scale flow,  and ${\bf \Omega}$ the vertical component of
the vorticity. The control parameter $\epsilon$ represents the reduced
Rayleigh number, while $\sigma$ characterizes the Prandtl number of the 
fluid. The parameter $g$ allows to more accurately reproduce the stability
properties of convection patterns\cite{CrGr}, and $g_m$  characterizes 
the coupling 
strength between the order parameter $\psi$ and the vorticity ${\bf \Omega}$. The
phenomenological parameter $c$ is introduced to describe the local dissipation 
of the vorticity (e.g. due to friction at the bottom of the convection
cell)\cite{CrGr,CrHoh}. Thus, Eq.~(\ref{SHE}) describes the dynamics of
the order parameter $\psi$, while Eq.~(\ref{Omega}), using the definition of the
vorticity (~\ref{vel}), represents the coupling of the large-scale flow field
${\bf u}$ and the order parameter. For $g = 1$ and $g_m = 0$
Eqs.~(\ref{SHE})-(\ref{vel}) reduce to the Swift-Hohenberg equation (\ref{she}).
We solved Eqs.~(\ref{SHE})-(\ref{vel}) numerically in a domain of
$256 \times 256$ collocation points using a pseudo-spectral method
based on the Fast Fourier Transform. The physical
domain size was typically restricted to $150 \times 150$.
Circular boundary conditions were enforced by ramping $\epsilon$ towards
negative values at distances from the center $r > R_{\mbox{max}} = 55$.

Essential features of SDC include spontaneous spiral creation, quasi-stationary 
spiral rotation, spiral core instability, and eventual spiral destruction 
by other spirals. Cross and Tu\cite{CrTu} showed that persistent spiral 
rotation is caused by the wavenumber frustration between the values selected
by the spiral core and by the environment (most typically, roll dislocations).
It is noteworthy that large vorticity generated in the spiral core plays only
the secondary role in this mechanism since the asymptotic wavenumber 
selected by the core does not depend on the  Prandtl number and coupling to 
the mean flow (parameter $g_m$ in Eq.(\ref{Omega})). However,
vorticity plays a major role in wavenumber selection by the roll dislocation.
It is easy to see that the dislocation in the roll pattern generates a
vortex pair which drives the dislocation towards the half-plane with greater
wavenumber (see Fig.1 where the order parameter for individual
spiral terminated by a dislocation is shown together with corresponding 
mean flow vorticity). At small $g_m\epsilon$ the dislocation
climbing speed due to vorticity is $v\propto g_m\epsilon$. The dislocation
is stationary when this mean-flow drift is balanced by the ordinary
climbing velocity $v_c\propto(1-q_d)^{3/2}$\cite{SigZip1}, therefore 
the selected wavenumber 
$q_{d0}\propto 1-\alpha (g_m\epsilon)^{2/3}$ ($\alpha$ is a constant depending
on other parameters $g,\sigma,c$).
In a stationary rotating spiral, wavenumber $q_d$ near the dislocation is 
larger than $q_{d0}$ so the dislocation moves around the core of the spiral 
with an angular velocity coinciding with the frequency of the core
rotation. It is easy to see from the phase diffusion equation (see \cite{CrTu})
that the frequency of the spiral rotation is inversely proportional to the 
distance from the core to the dislocation. Detailed measurements of individual
spiral rotation rate confirms this prediction\cite{boden}.  

At large supercriticality, a novel instability of the spiral core was recently 
observed\cite{PlBod,AssSt}. The core exhibits high-frequency oscillations
(with a period of several vertical diffusion times as compared with a period 
of overall spiral rotation of a few hundred vertical diffusion times). 
Using phase approximation (see \cite{CrHoh,CrTu}) it is easy to see that 
this instability is caused by
the large vorticity generated near the core of the spiral. Indeed, in the 
phase approximation, outside the core of the spiral the phase is described
by 
\begin{equation}
\theta_t+\mbox{\bf U}\cdot\nabla \theta=\tau(q)^{-1}\nabla\cdot [\mbox{\bf q} B(q)]
\label{phase}
\end{equation}
where $\tau(q),\ B(q)$ are known functions of the wavenumber 
$\mbox{\bf q}=\nabla\theta$
($B(1)=0,\ B(q_d)<0$ for SHE),see \cite{CrHoh}.  
For stationary rotating $n$-armed spiral, $\theta=\int q(r)dr+n\phi-\omega t$,
and therefore local wavenumber $q(r)$ obeys the following ODE,
\begin{equation}
\omega = -\frac{\Gamma}{r^2}-\tau(q)^{-1}\frac{d}{dr}[q B(q)],
\label{phase1}
\end{equation}
where $\Gamma$ is a total circulation of the mean flow around the spiral core.  
It is easy to see that for $1-q_d\ll 1$ the wavenumber 
$q(r)\approx 1-A_1r-A_2n\Gamma/r^2$ where $A_{1,2}$ are constants depending on 
$\tau$ and $B$.  The wavenumber slowly rises towards the core and then 
rapidly decreases in its near vicinity. Of course, this expression is only valid
outside the core of the dislocation ($r>1$), however it illustrates the 
tendency for 
the core vortex to locally unwind the spiral. If $\Gamma$ gets large, the 
wavenumber near the core goes outside the stability balloon and the spiral 
core becomes unstable. Visually it appears as that the core rotates against the
overall spiral rotation (see a series of snapshots in Fig.2 and computer movie 
at http://inls.ucsd.edu/$\sim$lev/conv/). This behavior is qualitatively
similar to observed in recent experiment by Plapp and
Bodenschatz\cite{PlBod}. The threshold of this core 
instability depends on both $\epsilon$ and $g_m$ as well as the 
topological charge $n$, see Fig.3. 

For yet larger values of supercriticality, a transition to hexagons in the 
core of the spiral occurs, see Fig.4. Both up- and down-flow hexagons
are generated simultaneously near the core of the spiral. The existence
of stable hexagons in systems without quadratic nonlinearity such as 
Swift-Hohenberg which preserves the symmetry $\psi\to-\psi$, was 
attributed to the spontaneous excitation of the zero mode by Dewell et al
\cite{Dew}. It should be noted that within the framework of
pure variational SHE (\ref{she}) the hexagonal state can only be metastable, so
a domain of roll will always expand towards a domain of hexagons.
In case of SHE coupled with mean flow, the zero mode is generated
and stabilized near the core by the mean flow which unwinds the spiral
and drives the wavenumber towards zero. This zero mode gives rise to a
proliferation of hexagons near the core of the spiral.
This transition was observed experimentally in \cite{AssSt,AAST} (see Fig.5).

\section{Spirals in a Faraday system}

Parametric instability of a flat fluid surface subject to vertical 
oscillations remains a popular experimental tool for various pattern 
formation phenomena. Squares, rolls, hexagons and higher-order 
quasi-patterns have been observed in different regions of parameters
(viscosity, driving frequency, and depth of the fluid layer). 
Phase diagram of different patterns has been recently computed by
Chen and Vi\~{n}als based on the Navier-Stokes equations\cite{ChVin}.
In the region of large viscosity and large $h/\lambda$,
straight rolls are a stable pattern. However, as in case of 
Rayleigh-B\'{e}nard convection, stable rotating spirals have been 
recently observed in the same parameter region\cite{EdFauve,KKRT}.
In this section we 
discuss the origin of the spiral rotation within the order-parameter 
model for the parametric instability,
\begin{equation}
\frac{\partial \psi }{\partial t}=\gamma \psi^*-\nu \psi
-(1+i\alpha )|\psi |^2\psi +i\kappa (\nabla ^2+1)\psi -({\bf u}\cdot
\nabla )\psi  \label{model}
\end{equation}
Here $\psi$ is a complex amplitude of surface oscillations at
the parametric frequency $\omega_0$ (which is a half of the driving frequency),
$\gamma $ is a forcing magnitude, $\kappa $ is the dispersion
parameter, and ${\bf u}$ is the velocity of the mean flow.
Linear terms in this equation can be derived from the dispersion relation
for capillary waves under parametric excitation, expanded near $\omega
=\omega _0,\,q=1$. The nonlinear term cannot be derived
rigorously, as in Eq.(\ref{SHE}) for Rayleigh-B\'{e}nard convection, 
and has been added {\em ad hoc} to account for the stabilization of 
the parametric instability.
Imaginary part of the nonlinear coefficient $\alpha $ describes nonlinear
frequency shift (see also \cite{vinals}). 

The last term in the r.h.s.  of Eq.(\ref{model}) describes order parameter 
advection by the mean flow. Such mean flow was observed in experiments
\cite{KKRT} near the walls of the cavity. This flow is caused by the momentum 
transfer from dissipating capillary waves at the driving frequency which are 
generated by oscillating side walls. This flow is directed off the walls near 
the surface and due to incompressibility returns back to the walls near the bottom 
Velocity ${\bf u}$ in (\ref{model}) should be understood as an the average velocity 
over layer thickness weighted with
the vertical structure of waves. Since surface waves decay towards the bottom,
near-surface flow affects them stronger than near-bottom return flow and the net
${\bf u}$ is oriented towards the center of the cavity.

Equation (\ref{model}) with periodic boundary conditions was studied
numerically using pseudo-spectral split-step method with $256\times 256$
collocation points, domain size $d=200$ and integration time step 0.05.
To simulate waves in circular cavity, we ramped linear dissipation outside
the circle of radius $r_0=86$, i.e. $\nu =\nu _0,r<r_0$ and
$\nu =\nu _0(1+k\,(r-r_0)),r>r_0$, where $k$ varied between 0.5 and 1.0.
We assumed that the flow had radial direction and was azimuthally
symmetric, ${\bf u}=u(r)\hat{{\bf \mbox{r}}}$. We used the following profile
for flow velocity $u(r):$
\begin{equation}
u(r)= u_0 \exp [\xi (r-r_0)]  \label{u}
\end{equation}
For $\gamma>\nu$ trivial state $\psi=0$ is unstable with respect to
perturbations with wavenumbers near 1. Numerical simulations show that
at the nonlinear stage, these perturbations give rise to various cellular
patterns, including plane waves, targets and spirals. Without mean flow term
($u_0=0$), these patterns remain stationary even when nonlinear
coefficient in (\ref{model}) is complex. Nonlinear frequency shift $%
\propto\alpha$ only leads to deviation of the selected wavenumber from $%
q=1 $. [In systems with ordinary (non-parametric)
pattern-forming instabilities non-potential effects usually lead to wave
propagation.] However, when near-wall flow (\ref{u}) is introduced in 
(\ref{model}), standing waves comprising
targets and spirals begin to drift slowly toward the center.
Figure 6 shows rolls velocity as a function of $u_0$ and $\xi$. Note 
that $\xi \gg r_0^{-1}$ so the flow is absent in the bulk, still rolls are
moving throughout the integration domain. The phase velocity
of rolls grows linearly with $u_0$ as it should be expected. 

This phenomenology can be understood in terms of the phase diffusion equation
similar to Eq.(\ref{phase}) for Rayleigh-B\'{e}nard system. The important difference
is that here we have standing waves, and therefore two phases
$\theta_+$ and $\theta_-$ should be introduced. However, for large parametric 
forcing, the sum of phases $\theta_+ +\theta_-$ is slaved to the phase 
difference $\Theta=\theta_+ -\theta_-$. The equation for the latter reads 
\begin{equation}
\partial_t\Theta + \mbox{\bf U}\cdot\nabla\Theta=\tau^{-1}(q)\nabla\cdot[\mbox{\bf
q}B(q)],
\label{phase2}
\end{equation}
where $\mbox{\bf q}=\frac{1}{2}\nabla\Theta$ and $\tau(q),B(q)$ are functions of
the wavenumber $q$. For the spiral, $\theta_\pm=\pm(\int qdr+m\phi)-
\omega_\pm t$. Multiplying (\ref{phase2}) by $\tau(q)$ and integrating from 
$r=0$ to $R_{\mbox{max}}$ we obtain
\begin{equation}
\omega_s\int \tau(q)dr =-\int U(r)q\tau(q)dr + qB(q)|_{r=R_{\mbox{max}}},
\label{omega}
\end{equation}
where $\omega_s=\omega_--\omega_+$ is a frequency of spiral rotation.
For $q\approx 1$ the last term in (\ref{omega}) is small and the frequency 
of spiral rotation can be estimated as 
$\omega_s\propto\int U(r)dr\approx u_0\xi^{-1}$. This relation is in 
an agreement with numerical simulations (see Figure 6). 

The mechanism of spiral rotation which we described is based on the assumption
that the radial flow is produced by the rapidly decaying waves near the wall generated
by the oscillating meniscus.  If this hypothesis is correct, the phase velocity of 
waves should increase linearly with the magnitude of driving. It also should depend 
sensitively on the wall profile. Indeed, experiment\cite{KKRT} confirmed this 
prediction (see Figure 7). 

\section{Conclusions}
In this paper we briefly considered two different pattern-forming systems
within the framework of order-parameter models 
coupled with mean flow. It turned out that both for Raylerigh-B\'{e}nard
convection and Faraday system the mean flow plays a crucial role for
spiral dynamics. In the former, vortex pairs generated by roll
dislocations, move the selected wavenumber away from the value selected
by the spiral core and lead to wavenumber frustration. The frustration
in turn leads to the persistent spiral rotation. The strong vortex which 
is generated in the core of the spiral plays a minor role in the overall
spiral rotation, however it leads to the local spiral unwinding and 
eventually to the spiral core instability. If the wavenumber near the core 
reaches zero due to local spiral unwinding, zero mode is generated and 
leads to proliferation of up- and down-flow hexagons.

In the Faraday system, stable rotating spirals have been recently observed. 
We showed that rotation of these spirals is caused by the near-wall
radial flow which leads to the wavenumber frustration in the bulk. This
near-wall flow is generated by rapidly decaying meniscus waves at
the driving frequency. 

The results presented in this paper have been obtained in collaboration 
with I.Aranson, M.Assenheimer, and V.Steinberg (Rayleigh-B\'{e}nard convection) 
and S.V.Kiyashko, L.N.Korzinov, and M.I.Rabinovich (Faraday system).
%Author acknowledges support by the U.S. Department of Energy,
%grants DE-FG03-95ER14516, DE-FG03-96ER14592.

\newpage 
\begin{figure}
\centerline{\epsfxsize = 3in \epsffile{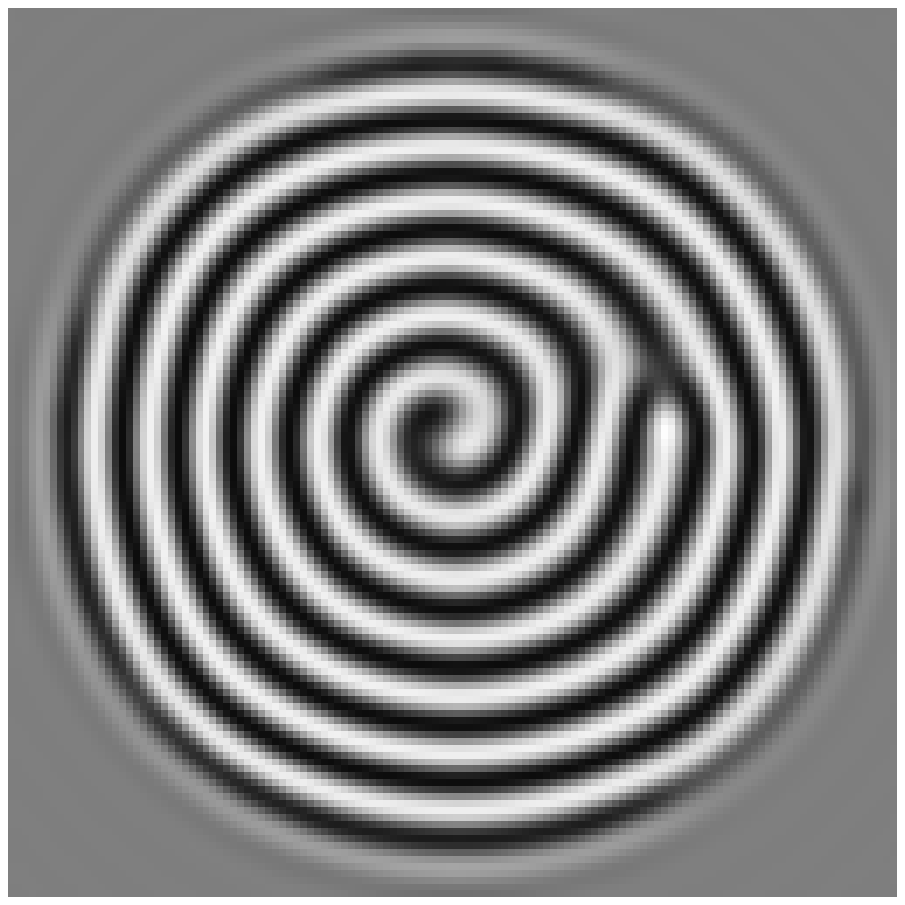},\epsfxsize = 3in \epsffile{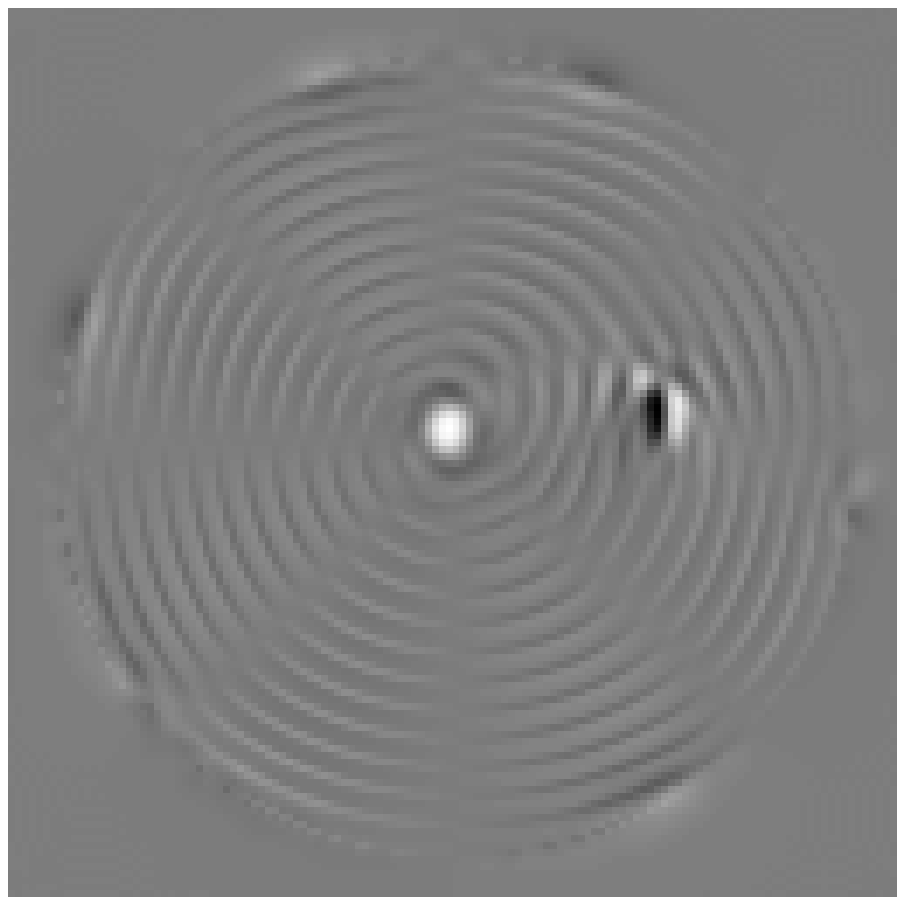}}
\caption{Snapshot of the order parameter $\psi$ (left) and vorticity $\Omega$ (right) for a
stationary rotating spiral terminated by a dislocation for Eqs.(\protect\ref{SHE})-
(\protect\ref{vel}) with $\epsilon=0.7,\ g=1,\ g_m=50,\ c^2=2,\ \sigma=1$.}
\end{figure}
\begin{figure}
\centerline{\epsfxsize = 8in \epsffile{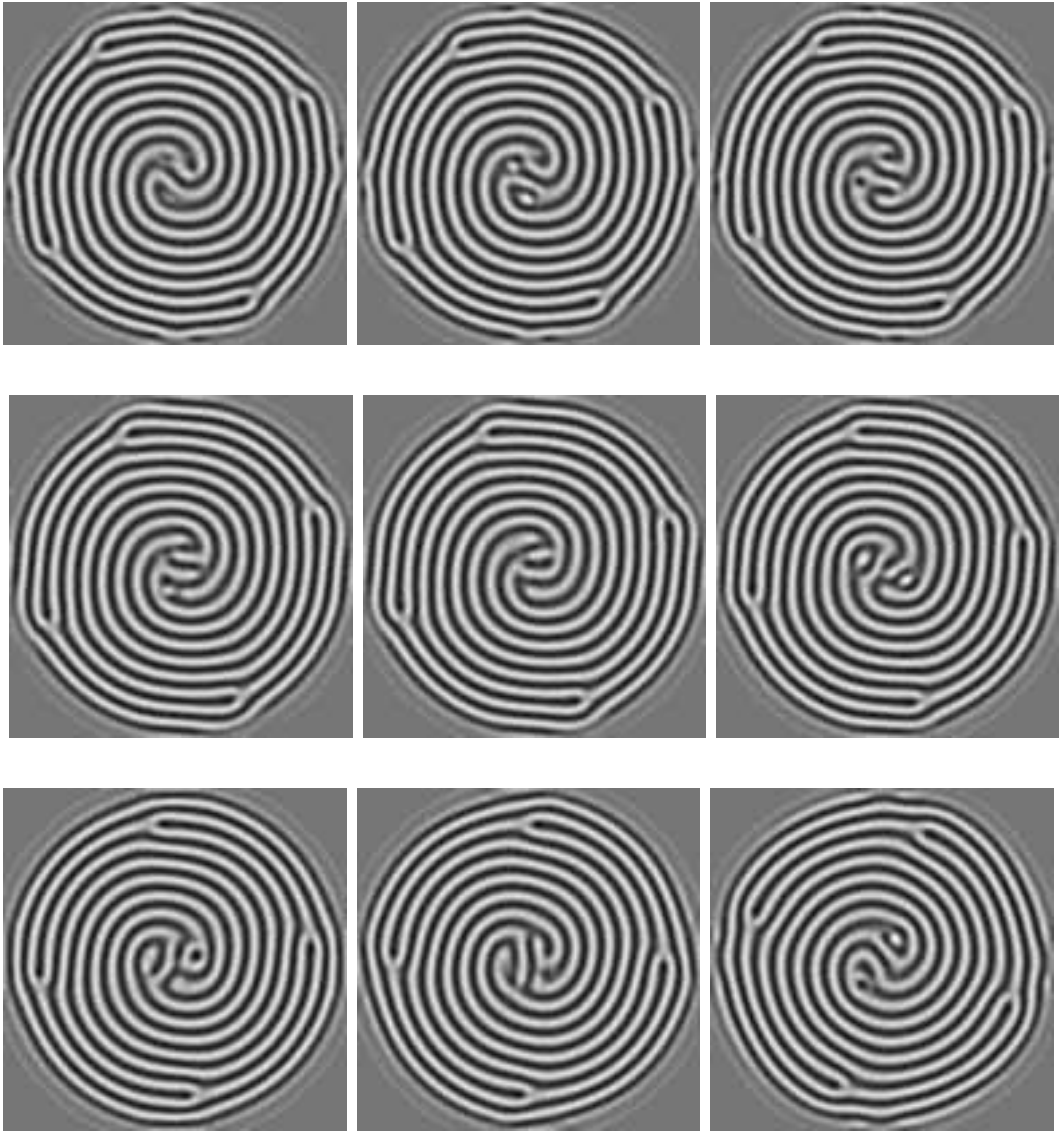}}
\caption{A sequence of snapshots of the order parameter for a 4-arm spiral with core
oscillations for Eqs.(\protect\ref{SHE})-(\protect\ref{vel}) with $\epsilon=0.5,\ g=1,\ g_m=50,
\ c^2=1,\ \sigma=1$.}
\end{figure}
\begin{figure}
\centerline{\epsfxsize = 4in \epsffile{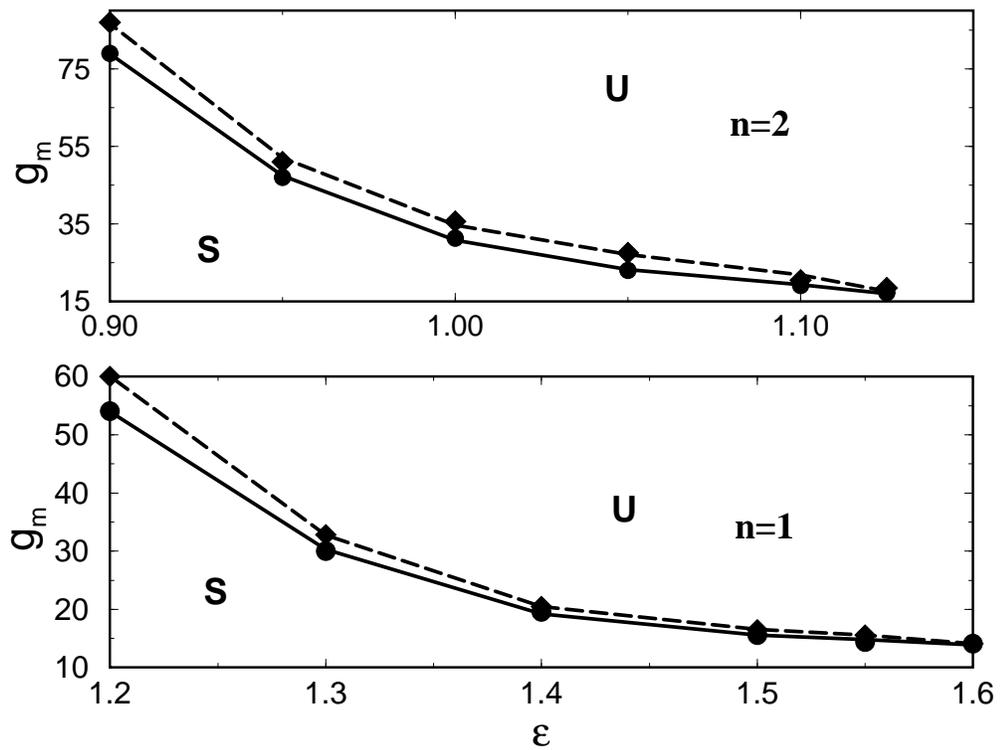}}
\caption{Stability diagram for one- (a) and two-armed (b) spirals for
$c^2=2$,$\sigma=1$ and $g=1$. As $\epsilon$ increases,  the spiral core becomes
unstable at the dashed line, regaining stability at the solid line as $\epsilon$
decreases.}
\end{figure}
\begin{figure}
\centerline{\epsfxsize = 4in \epsffile{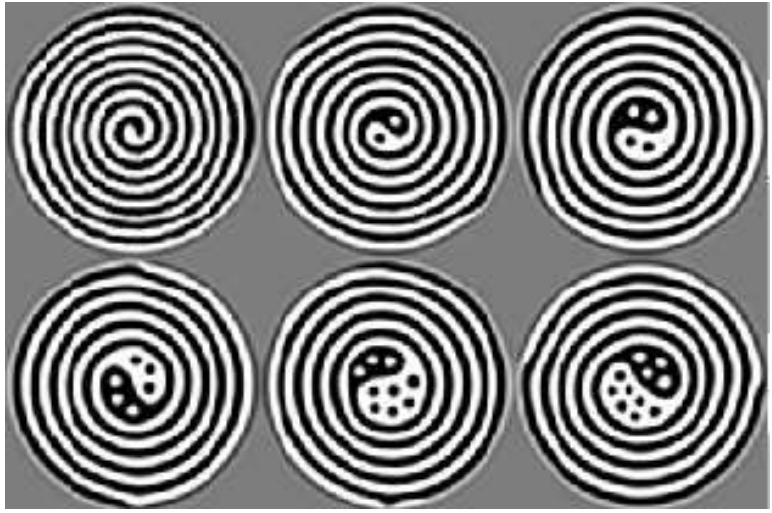}}
\caption{Hexagon nucleation at a spiral core obtained for Eqs.(\protect\ref{SHE})
-(\protect\ref{vel}) with $\epsilon=1.9, g=0.75, g_m=10, c^2=2$ and $\sigma=1$. 
Snapshots taken at $t = 10,110,470,650,340,2350$.}
\end{figure}
\begin{figure}
\centerline{\epsfxsize = 4in \epsffile{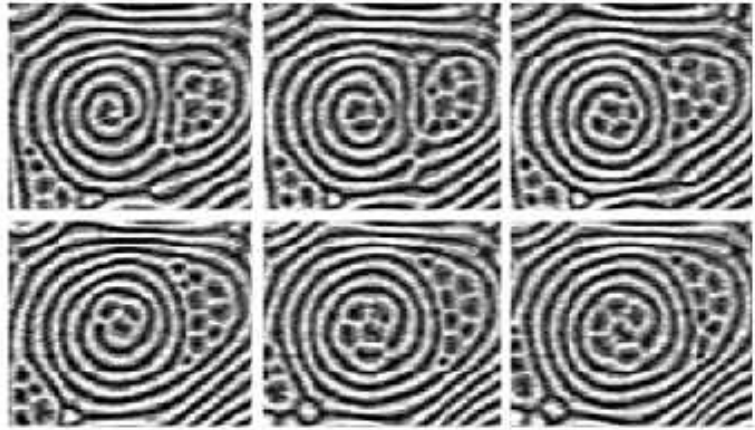}}
\caption{Hexagon nucleation at a spiral core in experiment with
$\epsilon = 3.19$, $\sigma = 4.5$ and time delay
between frames $\Delta t = 3.6$, $3.6$, $22.7$, $18.0$, $10.7$ $\tau_v$.}
\end{figure}
\begin{figure}
\centerline{\epsfxsize = 4in \epsffile{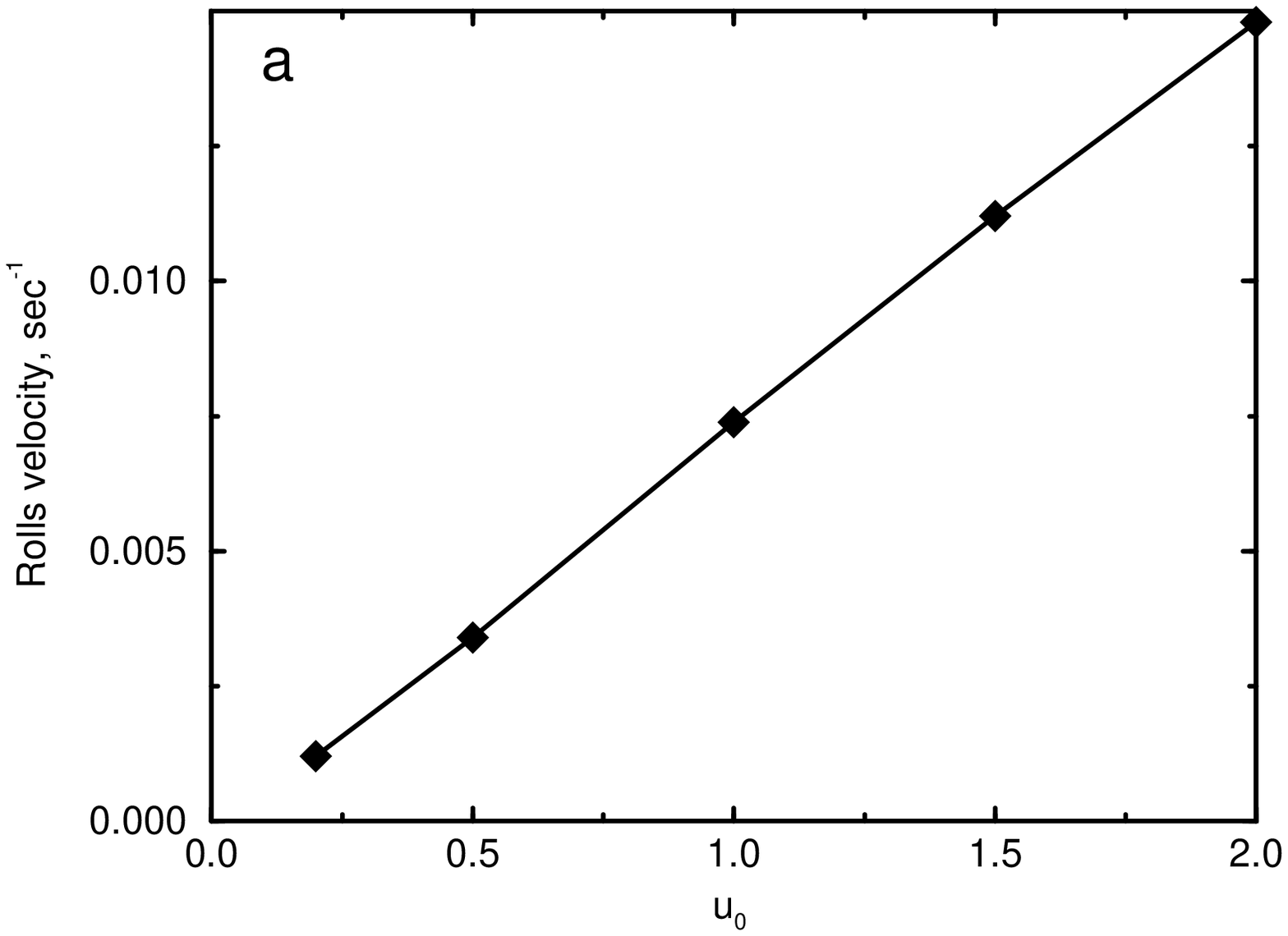}}
\centerline{\epsfxsize = 4in \epsffile{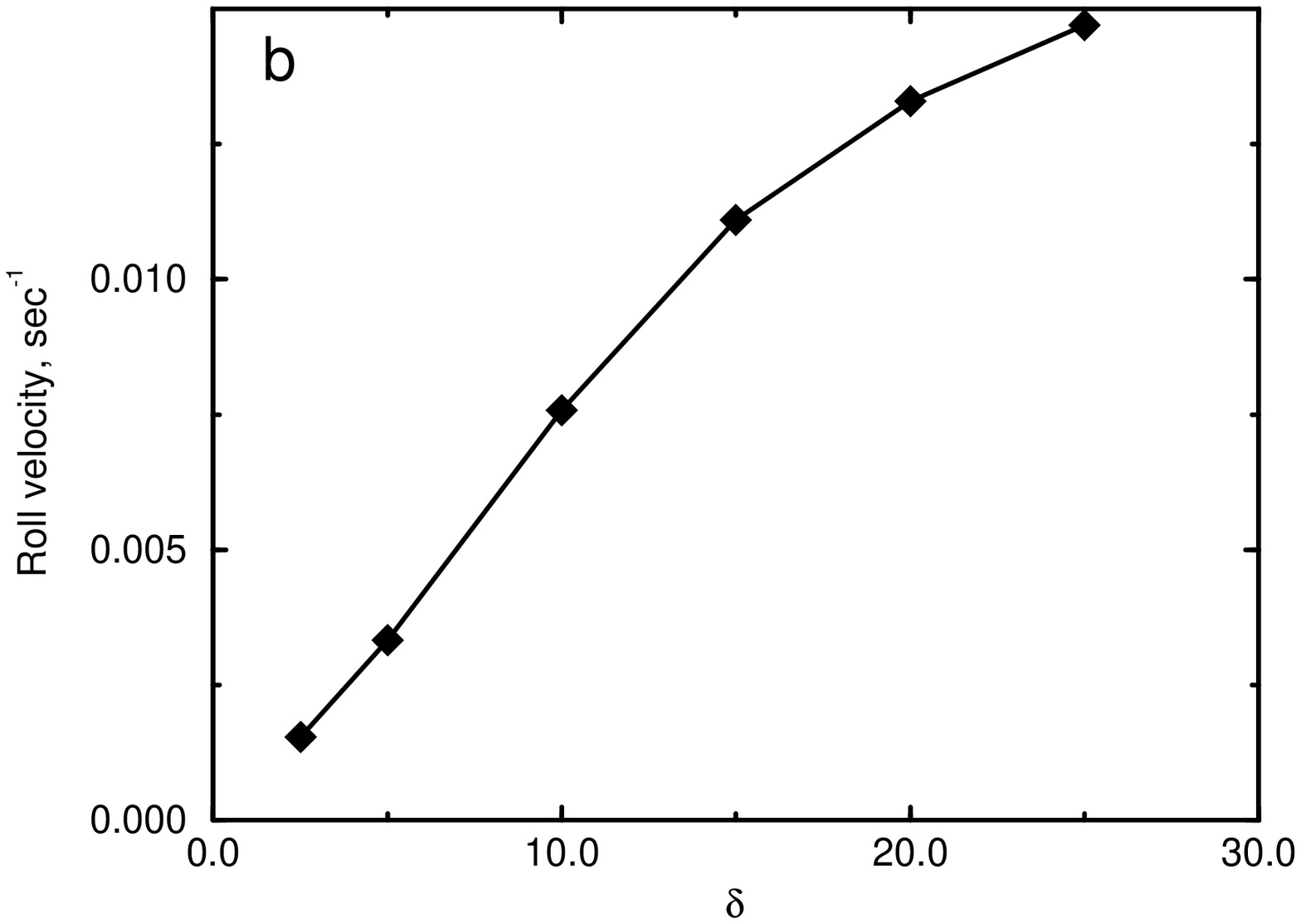}}
\caption{Phase velocity of waves as a function of near-wall flow magnitude $u_0$
({\em a}) and inverse scale $\xi $ ({\em b}) from numerical simulations of
(\protect\ref{model}).  Parameters of simulations: $\gamma =1.0, \nu =0.5, \alpha =0.0, \kappa
=1.0, k_0=1, \xi =0.1.$ In {\em a}, $\xi=0.1$, in {\em b},
$u_0=1.0$.}
\end{figure}
\begin{figure}
\centerline{\epsfxsize = 4in \epsffile{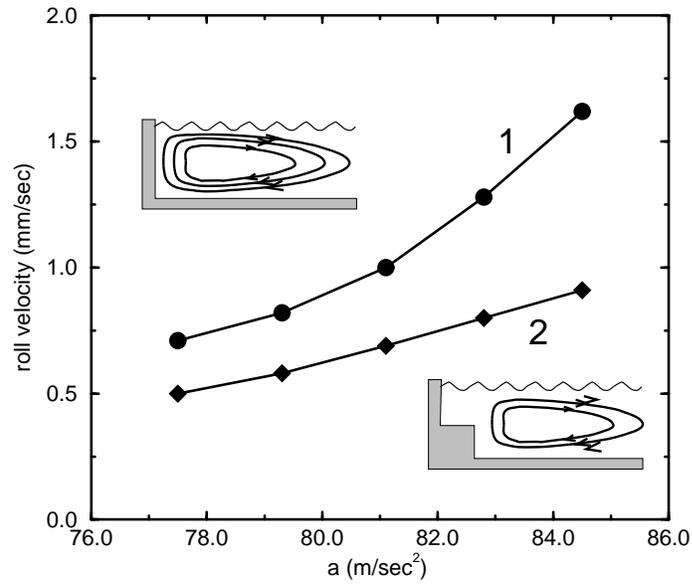}}
\caption{Velocity of wave drift (measured by nodes displacement) as a function
of magnitude of vertical
acceleration for two different profiles of side walls. {\em Inset:} Sketches
of the vertical profiles of the side wall and the structure of the shear flow.}
\end{figure}
\end{document}